\documentclass[aps,showpacs,preprint]{revtex4}
\usepackage{graphicx}
\usepackage{epsfig}
\DeclareGraphicsExtensions{.eps}

\begin{document}
\title{Seniority in quantum many-body systems}
\author{P.~Van~Isacker}
\affiliation{
Grand Acc\'el\'erateur National d'Ions Lourds,
CEA/DSM--CNRS/IN2P3, B.P.~55027, F-14076 Caen Cedex 5, France}

\begin{abstract}
The use of the seniority quantum number in many-body systems is reviewed.
A brief summary is given of its introduction by Racah
in the context of atomic spectroscopy.
Several extensions of Racah's original idea are discussed:
seniority for identical nucleons in a single-$j$ shell,
its extension to the case of many, non-degenerate $j$ shells
and to systems with neutrons and protons.
To illustrate its usefulness to this day,
a recent application of seniority is presented
in Bose--Einstein condensates of atoms with spin.
\end{abstract}

\maketitle

\section{Racah's seniority number}
\label{s_birth}
The seniority quantum number was introduced by Racah
for the classification of electrons in an atomic $\ell^n$ configuration~\cite{Racah43}.
He assumed a spin-independent interaction $\hat V$
between the electrons with the property
\begin{equation}
\langle\ell^2;LM_L|\hat V|\ell^2;LM_L\rangle=
g(2\ell+1)\delta_{L0},
\label{e_pairl1}
\end{equation}
that is, there is no interaction
unless the two electrons' orbital angular momenta $\ell$
are coupled to a combined angular momentum of $L=0$.
Racah was able to derive
a closed formula for the interaction energy among $n$ electrons
and to prove that any eigenstate of the interaction~(\ref{e_pairl1})
is characterized by a `seniority number' $\upsilon$,
a quantum number additional to the total orbital angular momentum $L$,
the total spin $S$
and the number of electrons $n$.
He also showed that $\upsilon$ corresponds
to the number of electrons that are not in pairs coupled to $L=0$~\cite{note1}.
Racah's original definition of seniority
made use of coefficients of fractional parentage.
He later noted that simplifications arose
through the use of group theory~\cite{Racah49}.
Seniority turned out to be a label
associated with the orthogonal algebra ${\rm SO}(2\ell+1)$
in the classification
\begin{equation}
{\rm U}(4\ell+2)\supset\Bigl({\rm U}(2\ell+1)\supset{\rm SO}(2\ell+1)\supset\cdots\supset{\rm SO}(3)\Bigr)
\otimes{\rm SU}_S(2),
\label{e_clasl}
\end{equation}
where the dots indicate intermediate algebras, if any exist.
The number of states available to a single electron
in an $\ell$ orbit is $4\ell+2$.
All states of the $\ell^n$ configuration
therefore belong to the totally antisymmetric irreducible representation (IR) $[1^n]$ of ${\rm U}(4\ell+2)$.
Furthermore, the natural scheme for electrons in an atom is $LS$ coupling
which corresponds to the reduction ${\rm U}(4\ell+2)\supset{\rm U}(2\ell+1)\otimes{\rm SU}_S(2)$,
where the orbital degrees of freedom are contained in ${\rm U}(2\ell+1)$
and the spin degrees of freedom in ${\rm SU}_S(2)$.
For any value of $\ell$ the unitary algebra ${\rm U}(2\ell+1)$
contains the orthogonal subalgebra ${\rm SO}(2\ell+1)$
which in turn contains SO(3),
associated with the total orbital angular momentum $L$.

The group-theoretical classification~(\ref{e_clasl})
allowed Racah to derive a number of important results
in the theory of complex atomic spectra.
The pairing force~(\ref{e_pairl1}), however, is a poor approximation
to the Coulomb interaction between electrons
and for a more physically relevant application of seniority
we have to turn to nuclei.

\section{Seniority in a single $j$ shell}
\label{s_single-j}
The discussion of seniority in atoms and in nuclei differs in two aspects:
(i) $LS$ coupling is a good first-order approximation in atoms
while in nuclei it is rather $jj$ coupling
and (ii) electrons are identical particles
while nucleons come in two kinds, neutrons and protons.
Let us postpone the discussion of the second complication until Sect.~\ref{s_isospin}
and concentrate in this section on the case of identical nucleons
(either all neutrons or all protons).
We impose the additional restriction that the identical nucleons
are confined to a single-$j$ shell,
deferring the discussion of the many-$j$ case to Sect.~\ref{s_many-j}.

It turns out that a pairing force of the type
\begin{equation}
\langle j^2;JM_J|\hat V|j^2;JM_J\rangle=
-g(2j+1)\delta_{J0},
\label{e_pairj1me}
\end{equation}
is a reasonable first-order approximation to the strong interaction
between identical nucleons.
In Eq.~(\ref{e_pairj1me}) $j$ is the total (orbital+spin) angular momentum of a single nucleon
and $J$ results from the coupling of two of them.
Since the pairing property now refers to the total $j$ of the nucleons,
there is no need for a separate treatment
of orbital and spin degrees of freedom as in Eq.~(\ref{e_clasl}),
and the classification becomes in fact simpler:
\begin{equation}
\begin{array}{ccccccc}
{\rm U}(2j+1)&\supset&{\rm Sp}(2j+1)&\supset&\cdots&\supset&{\rm SO}(3)\\
\downarrow&&\downarrow&&&&\downarrow\\[0mm]
[1^n]&&[1^\upsilon]&&&&J
\end{array}.
\label{e_clasj1}
\end{equation}
Seniority is associated with the (unitary) symplectic algebra ${\rm Sp}(2j+1)$
which replaces the orthogonal algebra ${\rm SO}(2\ell+1)$
of the atomic case.
Since the nucleons are identical,
all states of the $j^n$ configuration belong to
the totally antisymmetric IR $[1^n]$ of ${\rm U}(2j+1)$.
The IRs of ${\rm Sp}(2j+1)$
therefore must be totally antisymmetric of the type $[1^\upsilon]$.
The allowed values of seniority are $\upsilon=n,n-2,\dots,1$ or 0.
The angular momentum content for a given seniority $\upsilon$
can also be worked out~\cite{Wybourne70}
but no simple general rule is available.

An alternative, simpler definition of seniority can be given
which relies on the existence of an SU(2) symmetry
of the pairing hamiltonian~\cite{Kerman61,Helmers61}.
In second quantization
the pairing interaction~(\ref{e_pairj1me}) is written as
\begin{equation}
\hat V=-g\hat S^j_+\hat S^j_-,
\label{e_pairj1}
\end{equation}
with
\begin{equation}
\hat S^j_+=
{\frac 1 2}\sqrt{2j+1}\,
(a_j^\dag\times a_j^\dag)^{(0)}_0,
\qquad
\hat S^j_-=\left(\hat S^j_+\right)^\dag,
\end{equation}
where $a_{jm_j}^\dag$
creates a nucleon in the shell $j$ with projection $m_j$.
The commutator of $\hat S^j_+$ and $\hat S^j_-$
leads to the operator
$[\hat S^j_+,\hat S^j_-]=(2\hat n_j-2j-1)/2\equiv2\hat S^j_z$,
which thus equals, up to a constant, the number operator $\hat n_j$. 
Since the three operators $\{\hat S^j_z,\hat S^j_\pm\}$ close under commutation,
$[\hat S^j_z,\hat S^j_\pm]=\pm \hat S^j_\pm$
and $[\hat S^j_+,\hat S^j_-]=2\hat S^j_z$,
they form an SU(2) algebra,
referred to as the quasi-spin algebra.

This algebraic structure allows an analytical solution of the pairing hamiltonian.
From the commutation relations it follows that
$\hat S^j_+\hat S^j_-=(\hat S^j)^2-(\hat S^j_z)^2+\hat S^j_z$,
which shows that the pairing hamiltonian
can be written as a combination of Casimir operators
belonging to SU(2) and ${\rm SO}(2)\equiv\{\hat S^j_z\}$.
The associated eigenvalue problem can be solved instantly,
yielding the energy expression $-g[S(S+1)-M_S(M_S-1)]$.
The quantum numbers $S$ and $M_S$
can be put in relation to the seniority $\upsilon$
and the nucleon number $n$,
$S=(2j-2\upsilon+1)/4$ and $M_S=(2n-2j-1)/4$,
leading to the energy expression
$-g(n-\upsilon)(2j-n-\upsilon+3)/4$.
This coincides with the original expression given by Racah,
Eq.~(50) of Ref.~\cite{Racah43},
after the replacement of the degeneracy in $LS$ coupling, $4\ell+2$,
by the degeneracy in $jj$ coupling, $2j+1$.

While this analysis shows
that the eigenstates of a pairing interaction carry good seniority,
it does not answer the question
what are the necessary and sufficient conditions
for a general interaction to conserve seniority.
Let us specify a rotationally invariant two-body interaction $\hat V$
by the matrix elements
$\nu_J\equiv\langle j^2;JM_J|\hat V|j^2;JM_J\rangle$
with $J=0,2,\dots,2j-1$.
The necessary and sufficient conditions
for the conservation of seniority can then be written as
\begin{equation}
\sum_{J=2}^{2j-1}\sqrt{2J+1}
\left(\delta_{J I}+2\sqrt{(2J+1)(2I+1)}
\Bigl\{\begin{array}{ccc}j&j&J\\ j&j&I\end{array}\Bigr\}
-\frac{4\sqrt{(2J+1)(2I+1)}}{(2j-1)(2j+1)}
\right)\nu_J=0,
\label{e_condi}
\end{equation}
with $I=2,4,\dots,2j-1$,
and where the symbol between curly brackets is a Racah coefficient.
These conditions have been derived previously
in a variety of ways~\cite{Talmi93,Rowe01,Rosensteel03}.
Although (\ref{e_condi}) determines all constraints on the matrix elements $\nu_J$
by varying $I=2,4,\dots,2j-1$,
it does not tell us how many of those are independent.
This number turns out to be $\lfloor(2j-3)/6\rfloor$,
the number of independent seniority $\upsilon=3$ states~\cite{Ginocchio93}.
No condition on the matrix elements $\nu_J$
is obtained for $j=3/2$, 5/2 and 7/2,
one condition for $j=9/2$, 11/2 and 13/2, and so on.
As a result, identical nucleons in a single shell with $j\leq7/2$
conserve seniority for {\em any} interaction~\cite{Talmi93}.

Clearly, the conditions~(\ref{e_condi}) are much weaker
than the requirement that the interaction be of pairing character
but still many of the results of the quasi-spin formalism remain valid.
For instance, the ground state of an even--even nucleus
still can be written in the form~(\ref{e_cond1}).
The main restriction of the concept of seniority as defined so far,
concerns the fact that the nucleons are confined to a single-$j$ shell.
To lift this restriction, we turn to the generalization presented in the next section.

\section{Seniority in several $j$ shells}
\label{s_many-j}
The quasi-spin algebra can be generalized
to the case of several degenerate shells
(which we assume to be $s$ in number)
by making the substitutions
$\hat S^j_+\mapsto\hat S_+\equiv\sum_j\hat S^j_+$
and $2j+1\mapsto\sum_j(2j+1)$.
Therefore, if a semi-magic nucleus can be approximated
as a system of identical nucleons
interacting through a pairing force
and distributed over several degenerate shells,
the formulas of the quasi-spin formalism should apply. 
In particular, the ground states of even--even semi-magic nuclei
will have a `superfluid' structure of the form
\begin{equation}
\left(\hat S_+\right)^{n/2}|{\rm o}\rangle,
\label{e_cond1}
\end{equation}
where $|{\rm o}\rangle$ represents the vacuum
({\it i.e.}, the doubly-magic core nucleus).
The SU(2) quasi-spin solution of the pairing hamiltonian~(\ref{e_pairj1})
leads to several characteristic predictions:
a constant excitation energy (independent of $n$)
of the first-excited $2^+$ state in even--even isotopes,
the linear variation of two-nucleon separation energies as a function of $n$,
the odd--even staggering in nuclear binding energies,
the enhancement of two-nucleon transfer.

A more generally valid model is obtained
if one imposes the following condition on the hamiltonian:
\begin{equation}
[[\hat H,\hat S_+],\hat S_+]=
\Delta\left(\hat S_+\right)^2,
\label{e_gsen}
\end{equation}
where $\hat S_+$ creates the lowest two-nucleon eigenstate of $\hat H$
and $\Delta$ is a constant.
This condition of generalized seniority,
which was proposed by Talmi~\cite{Talmi71},
is much weaker than the assumption of a pairing interaction
and, in particular, it does not require
the commutator $[\hat S_+,\hat S_-]$
to yield (up to a constant) the number operator---a
property which is central to the quasi-spin formalism.
In spite of the absence of a closed algebraic structure,
it is still possible to compute the exact ground-state eigenvalue 
but hamiltonians satisfying~(\ref{e_gsen})
are no longer necessarily completely solvable.

\begin{figure}
\centering
\includegraphics[width=14cm]{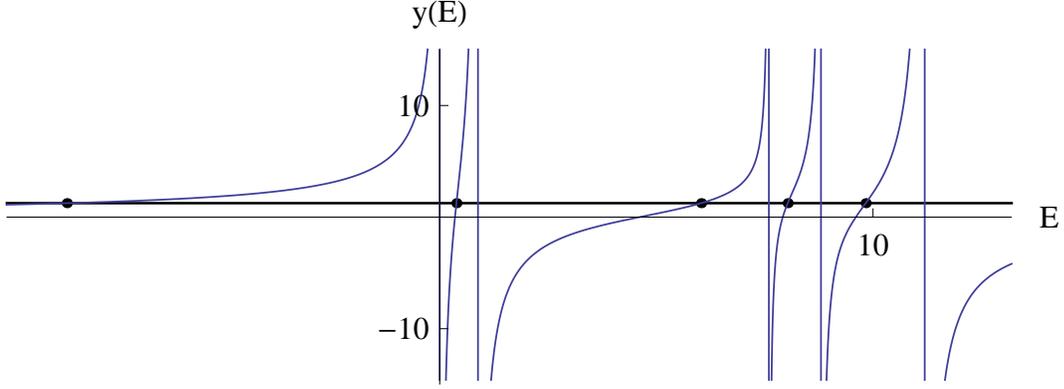}
\caption{
Graphical solution of the Richardson equation for $n=2$ fermions
distributed over $s=5$ single-particle orbits.
The sum $\sum_j\Omega_j/(2\epsilon_j-E)\equiv y(E)$
is plotted as a function of $E$;
the intersections of this curve with the line $y=1/g$ (dots)
then correspond to the solutions of the Richardson equation.}
\label{f_rich2}
\end{figure}

An exact method to solve the problem of identical nucleons
distributed over non-degenerate levels
interacting through a pairing force
was proposed a long time ago by Richardson~\cite{Richardson63}
based on the Bethe {\it ansatz}~\cite{Gaudin83}.
As an illustration of Richardson's approach,
we supplement the pairing interaction with a one-body term,
to obtain the following hamiltonian:
\begin{equation}
\hat H=
\sum_j\epsilon_j\hat n_j-g\hat S_+\hat S_-=
\sum_j\epsilon_j\hat n_j-g\sum_j\hat S^j_+\sum_{j'}\hat S^{j'}_-,
\label{e_pairjm1}
\end{equation}
where $\epsilon_j$ are single-particle energies.
The solvability of the hamiltonian~(\ref{e_pairjm1})
arises as a result of the symmetry
${\rm SU}(2)\otimes{\rm SU}(2)\otimes\cdots$
where each SU(2) algebra pertains to a specific $j$.
Whether the solution of~(\ref{e_pairjm1}) can be called superfluid 
depends on the differences $\epsilon_j-\epsilon_{j'}$
in relation to the strength $g$.
In all cases the solution is known in closed form
for all possible choices of $\epsilon_j$.
It is instructive to analyze first
the case of $n=2$ nucleons
because it gives insight
into the structure of the general problem.
The two-nucleon, $J=0$ eigenstates can be written as
$\hat S_+|{\rm o}\rangle=\sum_j x_j\hat S^j_+|{\rm o}\rangle$
with $x_j$ coefficients that are determined
from the eigenequation
$\hat H\hat S_+|{\rm o}\rangle=E\hat S_+|{\rm o}\rangle$
where $E$ is the unknown eigenenergy.
With some elementary manipulations
this can be converted into the secular equation
$2\epsilon_jx_j-g\sum_{j'}\Omega_{j'}x_{j'}=E x_j$,
with $\Omega_j=j+1/2$,
from where $x_j$ can be obtained 
up to a normalization constant,
$x_j\propto g/(2\epsilon_j-E)$.
The eigenenergy $E$ can be found
by substituting the solution for $x_j$
into the secular equation, leading to
\begin{equation}
\sum_j
{\frac{\Omega_j}{2\epsilon_j-E}}=
{\frac 1 g}.
\end{equation}
This equation can be solved graphically
which is done in Fig.~\ref{f_rich2}
for a particular choice of single-particle energies $\epsilon_j$
and degeneracies $\Omega_j$,
appropriate for the tin isotopes with $Z=50$ protons
and neutrons distributed over the 50--82 shell.
In the limit $g\rightarrow0$ of weak pairing,
the solutions $E\rightarrow2\epsilon_j$ are obtained,
as should be.
Of more interest is the limit of strong pairing,
$g\rightarrow+\infty$.
From the graphical solution we see that in this limit
there is one eigenstate of the pairing hamiltonian
which lies well below the other eigenstates
with approximately constant amplitudes $x_j$
since for that eigenstate $|E|\gg2|\epsilon_j|$.
Hence, in the limit of strong pairing
one finds a $J=0$ ground state
which can be approximated as
\begin{equation}
\hat S^{\rm c}_+|{\rm o}\rangle\approx
\sqrt{\frac{1}{\Omega}}\sum_j\hat S^j_+|{\rm o}\rangle,
\end{equation}
where $\Omega=\sum_j\Omega_j$.
Because of this property
this state is often referred to as the collective $S$ state,
in the sense that all single-particle orbits
contribute to its structure.

This result can be generalized to $n$ particles,
albeit that the general solution is more complex.
On the basis of the two-particle problem
one may propose, for an even number of particles $n$,
a ground state of the hamiltonian~(\ref{e_pairjm1}) 
of the form (up to a normalization constant)
\begin{equation}
\prod_{\alpha=1}^{n/2}
\left(\sum_j{\frac{1}{2\epsilon_j-E_\alpha}}\hat S^j_+\right)
|{\rm o}\rangle,
\label{e_gsrich}
\end{equation}
which is known as the Bethe {\it ansatz}~\cite{Gaudin83}.
Each pair in the product is defined
through coefficients $x_j=(2\epsilon_j-E_\alpha)^{-1}$
in terms of an energy $E_\alpha$
depending on $\alpha$ which labels the $n/2$ pairs.   
This product indeed turns out to be the ground state
provided the $E_\alpha$ are solutions of $n/2$ coupled, non-linear equations
\begin{equation}
\sum_j{\frac{\Omega_j}{2\epsilon_j-E_\alpha}}-
\sum_{\beta(\neq\alpha)}^{n/2}{\frac{2}{E_\beta-E_\alpha}}=
{\frac 1 g},
\qquad
\alpha=1,\dots,n/2,
\label{e_richeqs}
\end{equation}
known as the Richardson equations~\cite{Richardson63}.
Note the presence of a second term on the left-hand side
with differences of the unknowns $E_\beta-E_\alpha$ in the denominator,
which is absent in the two-particle case. 
In addition, the energy of the state~(\ref{e_gsrich})
is given by $\sum_\alpha E_\alpha$.
A characteristic feature of the Bethe {\it ansatz} is
that it no longer consists
of a superposition of {\it identical} pairs
since the coefficients $(2\epsilon_j-E_\alpha)^{-1}$
vary as $\alpha$ runs from 1 to $n/2$.
Richardson's model thus provides a solution
that covers all possible hamiltonians~(\ref{e_pairjm1}),
ranging from those with superfluid character
to those with little or no pairing correlations~\cite{Dukelsky04}.

An important remaining restriction
on the form of the pairing hamiltonian~(\ref{e_pairjm1})
is that it contains a single strength parameter $g$
whereas, in general, the interaction might depend on $j$ and $j'$,
leading to $s(s+1)/2$ strengths $g_{jj'}=g_{j'j}$.
In nuclei, often the assumption of a separable interaction is made
which, in the case of pairing, leads to strengths $g_{jj'}=gc_jc_{j'}$
in terms of $s$ parameters $c_j$.
This restriction leads to the following pairing hamiltonian:
\begin{equation}
\hat H=
\sum_j\epsilon_j\hat n_j-g\sum_{jj'}c_jc_{j'}\hat S^j_+\hat S^{j'}_-.
\label{e_pairjm2}
\end{equation}
As yet, no closed solution of the general hamiltonian~(\ref{e_pairjm2}) is known
but three solvable cases have been worked out:
\begin{enumerate}
\item
The strengths $c_j$ are constant (independent of $j$).
This case was discussed above.
\item
The single-particle energies $\epsilon_j$ are constant (independent of $j$).
The solution was given by Pan {\it et al.}~\cite{Pan98}
\item
There are two levels.
The solution was given by Balantekin and Pehlivan~\cite{Balantekin07}.
\end{enumerate}

\section{Seniority with neutrons and protons}
\label{s_isospin}
About ten years after its introduction by Racah,
seniority was adopted in nuclear physics
for the $jj$-coupling classification of nucleons
in a single-$j$ shell~\cite{Racah52,Flowers52}.
The main additional difficulty in nuclei
is that one deals with a system of neutrons and protons,
and hence the isospin $T$ of the nucleons should be taken into account.
The generalization of the classification~(\ref{e_clasj1})
for identical nucleons toward neutrons and protons reads as follows:
\begin{equation}
\begin{array}{ccccccccccc}
{\rm U}(4j+2)&\supset&
\Bigl({\rm U}(2j+1)&\supset&{\rm Sp}(2j+1)&\supset&\cdots&\supset&{\rm SO}(3)\Bigr)&
\otimes&{\rm SU}_T(2)\\
\downarrow&&\downarrow&&\downarrow&&&&\downarrow&&\downarrow\\[0mm]
[1^n]&&[h]&&[\sigma]&&&&J&&T
\end{array},
\label{e_clasj2}
\end{equation}
where $[h]$ and $[\sigma]$ are Young tableaux
associated with ${\rm U}(2j+1)$ and ${\rm Sp}(2j+1)$.
In general, $2j+1$ labels are needed to characterize
an IR of ${\rm U}(2j+1)$, $[h]=[h_1,h_2,\dots,h_{2j+1}]$,
and $j+1/2$ labels are needed for an IR of ${\rm Sp}(2j+1)$,
$[\sigma]=[\sigma_1,\sigma_2,\dots,\sigma_{j+1/2}]$.
To ensure overall antisymmetry under ${\rm U}(4j+2)$,
the Young tableaux of ${\rm U}(2j+1)$ and ${\rm U}_T(2)$ must be conjugate,
that is, one is obtained from the other
by interchanging rows and columns.
Since the Young tableau associated with ${\rm U}_T(2)$
is determined by the nucleon number $n$ and the total isospin $T$ as $[n/2+T,n/2-T]$,
the Young tableau of ${\rm U}(2j+1)$ must therefore be
\begin{equation}
[h]=[\overbrace{2,2,\dots,2}^{n/2-T},\overbrace{1,1,\dots,1}^{2T}].
\end{equation}
Since an IR of ${\rm U}(2j+1)$ has at most $2j+1$ labels,
it follows that $n/2+T\leq2j+1$.
Furthermore, all non-zero labels in $[\sigma]$ must be either 2 or 1
and the Young tableau of ${\rm Sp}(2j+1)$ must therefore be of the form
\begin{equation}
[\sigma]=[\overbrace{2,2,\dots,2}^{\upsilon/2-t},\overbrace{1,1,\dots,1}^{2t}].
\end{equation}
The IR of ${\rm Sp}(2j+1)$ is thus characterized by {\em two} labels:
the seniority $\upsilon$ and the `reduced isospin' $t$.
The former has the same interpretation as in the like-nucleon case
while the latter corresponds to the isospin of the nucleons
which are not in pairs coupled to $J=0$.

The group-theoretical analysis is considerably more complex here
than in the case of identical nucleons
and, in addition, for each value of $j$
one is faced with a different reduction problem
associated with ${\rm U}(2j+1)\supset{\rm Sp}(2j+1)\supset{\rm SO}(3)$.
It is therefore advantageous to go over
to a quasi-spin formulation of the problem
and, as was shown by Helmers~\cite{Helmers61},
this is possible for whatever value
of the intrinsic quantum number of the particles
(which is $t=1/2$ for nucleons).
If the pairing interaction is assumed to be isospin invariant,
it is the same in the three $T=1$ channels,
neutron--neutron, neutron--proton and proton--proton,
and Eq.~(\ref{e_pairj1}) can be generalized to
\begin{equation}
\hat V'=
-g\sum_\mu\hat S_{+,\mu}\hat S_{-,\mu}=
-g\hat S_+\cdot\hat S_-,
\label{e_pairj2}
\end{equation}
where the dot indicates a scalar product in isospin.
In terms of the nucleon creation operators $a^\dag_{jm_jtm_t}$,
which now carry also isospin indices (with $t=1/2$),
the pair operators are
\begin{equation}
\hat S_{+,\mu}=
{\frac 1 2}\sum_j\sqrt{2j+1}
(a_{jt}^\dag\times a_{jt}^\dag)^{(01)}_{0\mu},
\qquad
\hat S_{-,\mu}=
\left(\hat S_{+,\mu}\right)^\dag,
\label{e_pair2}
\end{equation}
where the coupling refers to angular momentum and to isospin.
The index $\mu$ (isospin projection) distinguishes
neutron--neutron ($\mu=+1$),
neutron--proton ($\mu=0$)
and proton--proton ($\mu=-1$) pairs.
There are thus three different pairs with $J=0$ and $T=1$
and they are related through the action
of the isospin raising and lowering operators $\hat T_\pm$.
By considering the commutation relations between the different operators,
a closed algebraic structure is obtained,
generated by the pair operators $\hat S_{\pm,\mu}$,
the number operator $\hat n$
and the isospin operators $\hat T_\pm$ and $\hat T_z$.
The quasi-spin algebra of neutrons and protons
in degenerate $j$ shells turns out to be SO(5),
by virtue of which the hamiltonian~(\ref{e_pairj2})
is analytically solvable~\cite{Hecht65,Ginocchio65}.

A further generalization is possible in $LS$ coupling.
For a neutron and a proton there exists
a different paired state with {\em parallel} spins.
The most general pairing interaction
for a system of neutrons and protons is therefore of the form
\begin{equation}
\hat V''=
-g\hat S_+\cdot\hat S_-
-g'\hat P_+\cdot\hat P_-,
\label{e_pairl2}
\end{equation}
where the pair operators are defined as
\begin{eqnarray}
\hat S_{+,\mu}=
\sqrt{\frac 1 2}\sum_\ell\sqrt{2\ell+1}
(a_{\ell st}^\dag\times a_{\ell st}^\dag)^{(001)}_{00\mu},
&\quad&
\hat S_{-,\mu}=
\left(\hat S_{+,\mu}\right)^\dag,
\nonumber\\
\hat P_{+,\mu}=
\sqrt{\frac 1 2}\sum_\ell\sqrt{2\ell+1}
(a_{\ell st}^\dag\times a_{\ell st}^\dag)^{(010)}_{0\mu0},
&\quad&
\hat P_{-,\mu}=
\left(\hat P_{+,\mu}\right)^\dag,
\label{e_pair3}
\end{eqnarray}
where $a_{\ell m_\ell sm_stm_t}^\dag$
creates a nucleon in the shell $\ell$ with projection $m_\ell$,
spin projection $m_s$ and isospin projection $m_t$.
The hamiltonian~(\ref{e_pairl2}) contains two parameters $g$ and $g'$,
the strengths of the isovector and isoscalar components
of the pairing interaction.
While in the previous case
the single strength parameter $g$
just defines an overall scale,
this is no longer true for a generalized pairing interaction
and different solutions are obtained for different ratios $g/g'$.

In general, the eigenproblem
associated with the interaction~(\ref{e_pairl2})
can only be solved numerically;
for specific choices of $g$ and $g'$
the solution of $\hat V''$
can be obtained analytically~\cite{Flowers64,Pang69}.
A closed algebraic structure is obtained,
formed by the pair operators~(\ref{e_pair3}),
their commutators,
the commutators of these among themselves,
and so on until closure is attained.
The quasi-spin algebra in this case turns out to be SO(8),
with 28 generators, consisting of
the pair operators $\hat S_{\pm,\mu}$ and $\hat P_{\pm,\mu}$,
the number operator $\hat n$,
the spin and isospin operators $\hat S_\mu$ and $\hat T_\mu$,
and the Gamow--Teller-like operator $\hat Y_{\mu\nu}$,
which is a vector in spin and isospin. 
The symmetry character of the hamiltonian~(\ref{e_pairl2})
is obtained by studying the subalgebras of SO(8).
Of relevance are the subalgebras
${\rm SO}_T(5)\equiv
\{\hat S_{\pm,\mu},\hat n,\hat T_\mu\}$,
${\rm SO}_T(3)\equiv\{\hat T_\mu\}$,
${\rm SO}_S(5)\equiv
\{\hat P_{\pm,\mu},\hat n,\hat S_\mu\}$,
${\rm SO}_S(3)\equiv\{\hat S_\mu\}$
and ${\rm SO}(6)\equiv
\{\hat S_\mu,\hat T_\mu,\hat Y_{\mu\nu}\}$,
which can be placed in the following lattice of algebras:
\begin{equation}
{\rm SO}(8)\supset
\left\{\begin{array}{c}
{\rm SO}_S(5)\otimes{\rm SO}_T(3)\\
{\rm SO}(6)\\
{\rm SO}_T(5)\otimes{\rm SO}_S(3)
\end{array}\right\}
\supset{\rm SO}_S(3)\otimes{\rm SO}_T(3).
\label{e_lattice}
\end{equation}
By use of the explicit form
of the generators of SO(8) and its subalgebras,
and their commutation relations~\cite{Pang69},
the following relations can be shown to hold:
\begin{eqnarray}
\hat S_+\cdot\hat S_-&=&
{\frac 1 2}\hat C_2[{\rm SO}_T(5)]-
{\frac 1 2}\hat C_2[{\rm SO}_T(3)]-
{\frac 1 8}(2\Omega-\hat n)(2\Omega-\hat n+6),
\nonumber\\
\hat S_+\cdot\hat S_-+
\hat P_+\cdot\hat P_-&=&
{\frac 1 2}\hat C_2[{\rm SO}(8)]-
{\frac 1 2}\hat C_2[{\rm SO}(6)]-
{\frac 1 8}(2\Omega-\hat n)(2\Omega-\hat n+12),
\nonumber\\
\hat P_+\cdot\hat P_-&=&
{\frac 1 2}\hat C_2[{\rm SO}_S(5)]-
{\frac 1 2}\hat C_2[{\rm SO}_S(3)]-
{\frac 1 8}(2\Omega-\hat n)(2\Omega-\hat n+6),
\end{eqnarray}
with $\Omega=\sum_\ell(2\ell+1)$
and where $\hat C_n[G]$ is the $n^{\rm th}$-order Casimir operator of the algebra $G$.
This shows that the interaction~(\ref{e_pairl2})
in the three cases
(i) $g=0$, (ii) $g'=0$ and (iii) $g=g'$,
can be written as a combination of Casimir operators
of algebras belonging to a chain of {\em nested} algebras
of the lattice~(\ref{e_lattice}).
They are thus the dynamical symmetries of the SO(8) model.

The nature of `SO(8) superfluidity' can be illustrated
in the specific example of the ground state of even--even $N=Z$ nuclei.
In the SO(6) limit of the SO(8) model
the exact ground-state solution can be written as~\cite{Dobes98}
\begin{equation}
\left(\hat S_+\cdot\hat S_+-\hat P_+\cdot\hat P_+\right)^{n/4}
|{\rm o}\rangle.
\label{e_quartet}
\end{equation}
This shows that the superfluid solution
acquires a {\em quartet} structure
in the sense that it reduces to a condensate of bosons
each of which corresponds to four nucleons.
Since the boson in~(\ref{e_quartet})
is a scalar in spin and isospin,
it can be thought of as an $\alpha$ particle;
its orbital character, however, might be different
from that of an actual $\alpha$ particle.
A quartet structure is also present
in the two SO(5) limits of the SO(8) model,
which yields a ground-state wave function of the type~(\ref{e_quartet})
with either the first or the second term suppressed.
A reasonable {\it ansatz}
for the $N=Z$ ground-state wave function
of the SO(8) pairing interaction~(\ref{e_pairl2})
with arbitrary strengths $g$ and $g'$ is therefore
\begin{equation}
\left(\cos\theta\;\hat S_+\cdot\hat S_+-\sin\theta\;\hat P_+\cdot\hat P_+\right)^{n/4}
|{\rm o}\rangle,
\label{e_trial}
\end{equation}
where $\theta$ is a parameter
that depends on the ratio $g/g'$.
The condensate~(\ref{e_trial}) of $\alpha$-like particles
provides an excellent approximation
to the $N=Z$ ground state
of the pairing hamiltonian~(\ref{e_pairl2})
for any combination of $g$ and $g'$~\cite{Dobes98}.
It should nevertheless be stressed that,
in the presence of both neutrons and protons
in the valence shell,
the pairing hamiltonian~(\ref{e_pairl2})
is {\em not} a good approximation
to a realistic shell-model hamiltonian
which contains an important quadrupole component.

These results can be generalized
to the case of several non-degenerate shells.
In fact, the Richardson equations~(\ref{e_richeqs})
are valid for the quasi-spin symmetry SU(2)
but they are known for any Lie algebra~\cite{Asorey02}.
Closed solutions have been obtained
for a system of neutron and protons with a pairing interaction
of pure isovector character
and of equal isovector and isoscalar strength,
based on the SO(5) and the SO(6) quasi-spin algebras,
respectively~\cite{Dukelsky05,Lerma07}.

\section{Bose--Einstein condensates of atoms with spin}
\label{s_bec}
In this section the concept of seniority
is illustrated with an application to the physics of cold atoms.
If atoms in a Bose--Einstein condensate (BEC)
are trapped by optical means~\cite{Stamper98},
their hyperfine spins (or spins) are not frozen in one particular direction
but are essentially free but for their mutual interactions.
As a result, the atoms do not behave as scalar particles
but each of the components of the spin
is involved in the formation of the BEC.
This raises interesting questions concerning the structure of the condensate
and how it depends on the spin-exchange interactions between the atoms.

Such questions were addressed
in a series of theoretical papers by Ho and co-workers~\cite{Ho98}
who obtained solutions based on a generating function method.
In the case of spin-1 atoms the problem of quantum spin mixing 
was analyzed by Law {\it et al.}~\cite{Law98}
who proposed an elegant solution based on algebraic methods.
It is shown here that an exact solution
is also available for the spin value $f=2$
(for any number of atoms $n$)
which allows the analytic determination
of the structure of the ground state of the condensate.
This was simultaneously and independently pointed out
in Refs.~\cite{Isacker07,Uchino08}.

We consider a one-component dilute gas of trapped bosonic atoms 
with arbitrary (integer) hyperfine spin $f$.
In second quantization the hamiltonian of this system
has a one-body and a two-body piece that can be written
as (in the notation of Ref.~\cite{Law98})
\begin{equation}
{\cal H}=
\sum_m\int
\hat\Psi_m^\dag
\left(-\frac{\nabla^2}{2M_{\rm a}}+V_{\rm trap}\right)\hat\Psi_m d^3x+
\sum_{m_i}\Omega_{m_1m_2m_3m_4}
\int\hat\Psi_{m_1}^\dag\hat\Psi_{m_2}^\dag\hat\Psi_{m_3}\hat\Psi_{m_4}d^3x,
\label{e_hambec}
\end{equation}
where $\hbar=1$, $M_{\rm a}$ is the mass of the atom,
and $\hat\Psi_m$ and $\hat\Psi_m^\dag$
are the atomic field annihilation and creation operators
associated with atoms in the hyperfine state $|fm\rangle$
with $m=-f,\dots,+f$,
the possible values of all summation indices in~(\ref{e_hambec}).
The trapping potential $V_{\rm trap}$
is assumed to be the same for all $2f+1$ components.
According to the assumptions outlined in Ref.~\cite{Law98},
the atomic field creation and annihilation operators at zero temperature
can be approximated by
$\hat\Psi_m^\dag\approx b_m^\dag\phi(\vec x)$,
$\hat\Psi_m\approx b_m\phi(\vec x)$,
$m=-f,\dots,+f$,
where $\phi(\vec x)$ is a single wave function (independent of $m$)
and $b_m$ and $b_m^\dag$
are annihilation and creation operators,
satisfying the usual boson commutation rules.
In this approximation the entire hamiltonian~(\ref{e_hambec})
can be rewritten as
\begin{equation}
{\cal H}\approx\hat H\equiv
\epsilon\,b^\dag\cdot\tilde b+
{\frac 1 2}\sum_F\nu_F(b^\dag\times b^\dag)^{(F)}\cdot
(\tilde b\times\tilde b)^{(F)},
\label{e_hamgen}
\end{equation}
where the coefficients $\epsilon$ and $\nu_F$
are related to those in the original hamiltonian~(\ref{e_hambec})
and with $\tilde b_m\equiv(-)^{f-m}b_{-m}$.

Exactly solvable hamiltonians with rotational or SO(3) invariance
are now found by the determination of all Lie algebras $G$
satisfying ${\rm U}(2f+1)\supset G\supset{\rm SO}(3)$.
The canonical reduction of U($2f+1$) is of the form
as encountered by Racah (see Sect.~\ref{s_birth}),
\begin{equation}
{\rm U}(2f+1)\supset{\rm SO}(2f+1)\supset{\rm SO}(3),
\label{e_chain}
\end{equation}
defining a class of solvable hamiltonians of the type
\begin{equation}
\hat H'=
a_1\hat C_1[{\rm U}(2f+1)]+
a_2\hat C_2[{\rm U}(2f+1)]+
b\,\hat C_2[{\rm SO}(2f+1)]+
c\,\hat C_2[{\rm SO}(3)],
\label{e_hamsol}
\end{equation}
where $a_1$, $a_2$, $b$, and $c$ are numerical coefficients.
The solvability properties of the original hamiltonian~(\ref{e_hamgen})
now follow from a simple counting argument.
For atoms with spin $f=1$
the solvable hamiltonian~(\ref{e_hamsol})
has three coefficients $a_1$, $a_2$, and $c$
[since SO($2f+1$)=SO(3)]
while the general hamiltonian~(\ref{e_hamgen})
also contains three coefficients $\epsilon$, $\nu_0$, and $\nu_2$.
(Note that the coupling of two spins to odd $F$
is not allowed in the approximation
of a common spatial wave function, so no $\nu_1$ term occurs.)
They can be put into one-to-one correspondence.
For atoms with spin $f=2$
both the solvable and the general hamiltonian contain four coefficients
($a_1$, $a_2$, $b$, and $c$
{\it versus} $\epsilon$, $\nu_0$, $\nu_2$, and $\nu_4$)
which also can be put into one-to-one correspondence.
Hence the general hamiltonian~(\ref{e_hamgen}) is solvable for $f\leq2$.
The same counting argument shows
that it is no longer solvable for $f>2$.

The case of interacting $f=1$ atoms was discussed by Law {\it et al.}~\cite{Law98}
who identified the existence of two possible condensate ground states:
one with all atoms aligned to maximum spin $F=n$
and a second with pairs of atoms coupled to $F=0$.
Whether the condensate is aligned or paired
depends on a single interaction parameter.
With the technique explained above,
the phase diagram for atoms with spin $f=2$
can also be derived.
The results are exact and valid for arbitrary $n$.
The entire spectrum is determined by the eigenvalue expression
together with the necessary branching rules.
In particular, the allowed values of total spin $F$ for a given seniority $\upsilon$
are derived from the ${\rm SO}(5)\supset{\rm SO}(3)$ branching rule
given by $F=2\tau,2\tau-2,2\tau-3,\dots,\tau+1,\tau$
with $\tau=\upsilon,\upsilon-3,\upsilon-6,\dots$ and $\tau\geq0$.

It is now possible to determine
all possible ground-state configurations of the condensate
and their quantum numbers $\upsilon_0$ and $F_0$~\cite{Isacker07}.
The character of the ground state
does not depend on the coefficients $a_i$
since the first two terms in the expression~(\ref{e_hamsol})
give a constant contribution to the energy of all states.
Although this contribution is dominant,
the spectrum-generating perturbation of the hamiltonian
is confined to the last two terms
and depends solely on the coefficients $b$ and $c$
which are related to the original interactions $\nu_F$
according to
$b=(-7\nu_0+10\nu_2-3\nu_4)/70$
and $c=(-\nu_2+\nu_4)/14$.
The phase diagram displays a richer structure than in the $f=1$ case.
There is an aligned phase
where the seniority is maximal, $\upsilon_0=n$,
and all spins are aligned, $F_0=2n$.
Secondly, there is a low-seniority (paired) and consequently low-spin phase.
For even $n$, this corresponds to $(\upsilon_0,F_0)=(0,0)$.
The aligned and paired phases
are also encountered for interacting $f=1$ atoms.
For $f=2$ a third phase occurs
characterized by high seniority ({\it i.e.}, unpaired) and low total spin,
$(\upsilon_0,F_0)=(n,2\delta)$ with $\delta=0$ or 1.

Since the hamiltonian~(\ref{e_hamsol}) is solvable for $f=2$,
all eigenstates, and in particular the three different ground states,
can be determined analytically.
The general expressions given by Chac\'on {\it et al.}~\cite{Chacon76} reduce to
\begin{eqnarray}
|\upsilon=n,F=M_F=2n\rangle&\propto&
\left(d_{+2}^\dag\right)^n|{\rm 0}\rangle,
\nonumber\\
|\upsilon=0,F=M_F=0\rangle&\propto&
\left(d^\dag\cdot d^\dag\right)^{n/2}|{\rm 0}\rangle,
\nonumber\\
|\upsilon=n,F=M_F=0\rangle&\propto&
\left((a^\dag\times a^\dag)^{(2)}\cdot a^\dag\right)^{n/3}|{\rm 0}\rangle,
\label{e_wave}
\end{eqnarray}
where the $f=2$ atoms are denoted as $d$ bosons.
In the second of these expressions it is assumed that $n$ is even
and in the third that $n=3k$;
other cases are obtained
by adding a single boson $d^\dag$ or a $d^\dag\cdot d^\dag$ pair.
The $a^\dag$ are the so-called traceless boson operators~\cite{Chacon76}
which are defined as
$a_m^\dag=d_m^\dag-d^\dag\cdot d^\dag(2\hat n+5)^{-1}\tilde d_m$
(see also Chapt.~8 of Ref.~\cite{Frank94}).
The wave functions~(\ref{e_wave}) are the {\em exact} finite-$n$ expressions
for the eigenstates of the hamiltonian~(\ref{e_hamsol}).
Since in the large-$n$ limit the traceless boson operators $a_m^\dag$
become identical to $d_m^\dag$,
one arrives at a simple interpretation of the three types of configurations:
(i) spin-aligned,
(ii) condensed into {\em pairs of atoms} coupled to $F=0$,
and (iii) condensed into {\em triplets of atoms} coupled to $F=0$.

In conclusion, the consideration of seniority is crucial in obtaining results
concerning Bose--Einstein condensates consisting of atoms with spin.
Since all eigenstates of interacting atoms with spin $f\leq2$ are known analytically,
this opens up the possibility to study the relaxation properties
of such condensates using their exact, macroscopic wave functions.
In addition, preliminary studies indicate that seniority
can be exploited even when $f>2$.
These problems are currently under investigation~\cite{Isackerun}.

This paper is dedicated to the memory of Marcos Moshinsky.
The two years I have spent in Mexico as a visitor
and the many hours with Marcos as a teacher,
were crucial to my formation as a physicist.
Without him I never could have written this paper.


\begin{thebibliography}{99}
\bibitem{note1}
This is a rather loose statement in words.
The precise definition of seniority
involves a property of vanishing $n$-to-$(n-2)$
coefficients of fractional parentage.
\end{thebibliography}
\end{document}